\documentclass[aps,twocolumn,superscriptaddress,showpacs]{revtex4-2}

\usepackage{graphicx}
\usepackage{color}
\usepackage{dcolumn}
\usepackage{bm}
\usepackage{here}
\usepackage{amsmath}
\usepackage{longtable}
\usepackage{hyperref}

\begin{document}

\preprint{}

\title{Microscopic dynamics of lithium diffusion 
in single crystal of the solid-state electrolyte La$_{2/3-x}$Li$_{3x}$TiO$_{3}$ ($x=0.13$)
studied by quasielastic neutron scattering
}

\author{Masato~Matsuura$^{*}$}
  \affiliation{Neutron Science and Technology Center, Comprehensive Research Organization for Science and Society (CROSS), Tokai, Ibaraki 319-1106, Japan}
  \email{m_matsuura@cross.or.jp}
\author{Yasuyuki Fujiwara}
  \affiliation{Faculty of Engineering, Shinshu University, Nagano 380-8553, Japan}
\author{Hiroki Moriwake}
  \affiliation{Nanostructures Research Laboratory, Japan Fine Ceramics Center, Nagoya 456-8587, Japan}
\author{Koji Ohara}
  \affiliation{Research and Utilization Division, Japan Synchrotron Radiation Research Institute (SPring-8/JASRI),  Sayo, Hyogo 679-5198, Japan}
\author{Yukinobu Kawakita}
  \affiliation{Materials and Life Science Division, J-PARC Center, JAEA, Tokai, Ibaraki 319-1195, Japan}
\date{\today}

\begin{abstract}
Quasi-elastic neutron scattering (QENS) measurements 
combined with first principles based molecular-dynamics calculations 
were conducted to study the dynamics of Li$^+$ ions
in a solid-state electrolyte La$_{2/3-x}$Li$_{3x}$TiO$_{3}$ (LLTO) with $x=0.13$.
By using a large $^{7}$Li-enriched single crystal sample,
a QENS signal was clearly observed along the three principal axes 
[110], [111], and [001] at a temperature ($T$) of 600~K.
Wave vector dependence of the line width of the QENS signal 
along each direction was explained well using the Chudley-Elliot model for 
jumps between the $A$ sites of the perovskite lattice
through the bottleneck square,
which was also supported by molecular dynamics calculations.
At $T=600$~K, the estimated self-diffusion coefficient of Li$^+$ ($D_{\rm Li}$) 
in the $ab$-plane [$D_{\rm Li}^{ab}=(6.8\pm0.5)\times10^{-6}~$cm$^2$/s] was slightly larger
than that along the $c$ axis [$D_{\rm Li}^{c}=(4.4\pm0.3)\times10^{-6}~$cm$^{2}$/s],
suggesting quasi-isotropic diffusion, that is, the three-dimensional diffusion of Li$^+$ ions.
The decrease in $D_{\rm Li}$ with decreasing $T$ was reasonably explained by a thermal activation process with
the activation energy determined from ionic-conductivity measurements. 
Furthermore, the estimated values of the self-diffusion coefficient of Li$^+$ ions 
are comparable to those in the sulfide-based Li$^+$ ion conductor,
Li$_{7}$P$_{3}$S$_{11}$, 
although its ionic conductivity is 10 times larger than that for LLTO. 
The obtained microscopic information on Li$^+$ diffusion in LLTO clarifies how to 
understand the Li conduction mechanism in LLTO and Li$_{7}$P$_{3}$S$_{11}$ in a unified manner and 
can provide a way to increase the Li$^+$ ionic conductivity in oxide-based solid electrolytes.
\end{abstract}


\maketitle

\section{Introduction}
An all-solid-state lithium-ion battery has been heavily investigated  
as a next-generation energy storage system \cite{Tarascon01,Armand08}, 
because both the energy density and safety of such 
all-solid-state batteries are expected to be drastically improved 
by replacing an organic solvent-based liquid electrolyte with a solid electrolyte.
Sulfide-based Li$^+$ ionic conductors, 
such as Li$_{10}$GeP$_{2}$S$_{12}$ (LGPS),
are considered to be promising candidates for a solid electrolyte
because of their extremely high Li$^+$ ionic conductivity ($\sigma_{\rm Li}$) ranging
approximately $10^{-2}$~S/cm at room temperature, 
which is comparable to the $\sigma_{\rm Li}$ of typical liquid electrolytes~\cite{Kamaya11}.
By contrast, $\sigma_{\rm Li}$ of oxide-based Li$^+$ ionic conductors
is more than one order of magnitude lower than that of  
sulfide-based Li$^+$ ionic conductors such as LGPS, Li$_3$PS$_4$, and Li$_7$P$_3$S$_{11}$
as summarized in Table~\ref{tb1}.
Because oxide-based Li$^+$ ionic conductors have an excellent stability in air,
it is highly desirable to develop oxide-based Li$^+$ ionic conductors
with a high $\sigma_{\rm Li}$. 

Among the many oxide-based Li$^+$ ionic conductors,
La$_{2/3-x}$Li$_{3x}$TiO$_{3}$ (LLTO) exhibits 
the highest $\sigma_{\rm Li}$ at room temperature~\cite{Latie84,Inaguma93,Inaguma96}.
Since LLTO poses a simple double-perovskite structure
[see Fig.~\ref{fig1}(a)], 
LLTO is considered an ideal system for studying 
the relationship between the Li$^+$ conduction mechanism 
and structural properties in solids.
Previous extensive structural studies on LLTO 
revealed the following key features on $\sigma_{\rm Li}$ 
\cite{Inaguma93, Fourquet96, Harada98, Ibarra00, Mazza02, Inaguma_06, Ohara10, Mori14}.
First, $\sigma_{\rm Li}$ is closely related to the La and defect contents at
the $A$ site in a double-perovskite lattice, 
in which the La-rich and La-poor layers are stacked alternately along the $c$ axis.  
Li$^+$ ions mainly conduct through the La-poor layer, 
when the Li concentration is low (or $x<0.08$)~\cite{Inaguma93,Ibarra00,Mazza02,Inaguma_06,Ohara10}.
The difference in the La occupancy between
the La-rich and La-poor layers decreases with the Li content. 
A nearly three-dimensional diffusion is expected 
for an Li-rich composition. 
Second, $\sigma_{\rm Li}$ strongly depends on the size of
the ``bottleneck'' square surrounded by oxygen ions.
The potential barrier of $\sigma_{\rm Li}$ is dominated
by the repulsion energy from oxygen ions~\cite{Inaguma_06,Mori14}.

Unfortunately, these findings are based on the time-averaged structure
and do not provide direct information on the dynamics of Li$^+$ ions.
However, neutron scattering can detect the motion of Li$^+$ ions
as quasielastic neutron scattering (QENS). 
When a neutron is scattered by a mobile Li$^+$ ion,
neutron energy is transferred to, or from, the Li$^+$ ions,
resulting in broadening of the elastic signal.
The wave vector ($Q$) dependence of the line-width of the QENS 
provides detailed information on the dynamical conduction path of Li$^+$
on a microscopic scale, such as the residence time ($\tau$) and jump vector ($l$).
Furthermore, if a single crystal is available, 
directional information of Li$^+$ jump can be obtained.
However, such a detailed QENS study using a single crystal 
has not been reported thus far 
perhaps owing to a lack of sufficiently large single crystals
and weak QENS signals from mobile Li$^+$ ions.

In this paper, 
we present a QENS study on the dynamics of Li$^+$ ions using a single crystal of LLTO 
and a state-of-the-art neutron backscattering spectrometer.
Combined with a molecular dynamics (MD) simulation using first-principles calculations,
the dynamical conduction path and direction dependent 
self-diffusion coefficient of Li$^+$ ions 
($D_{\rm Li}$) at a microscopic scale
have been directly extracted from QENS data.

\begin{table*}
\caption{\label{tb1}Comparison of ionic conductivity ($\sigma_{\rm Li}$), activation energy ($E_a$), 
and diffusion coefficient ($D_{\rm Li}$) for LGPS, Li$_3$PS$_4$, Li$_7$P$_3$S$_{11}$ (glass), and LLTO. 
$\sigma_{\rm Li}$ and $E_a$ are the values at room temperature (RT). 
$D_{\rm Li}$'s for LGPS and Li$_3$PS$_4$ are the reported values obtained
from pulsed field gradient nuclear magnetic resonance measurements,
whereas $D_{\rm Li}$'s for Li$_7$P$_{3}$S$_{11}$ and LLTO are obtained from QENS measurements.
$^{*}D_{\rm Li}$ at $T=473$~K for Li$_3$PS$_4$ and LLTO are estimated 
using the Arrhenius law with $E_{a}$.}
\begin{ruledtabular}
\begin{tabular}{c|c|c|c}
Systems & $\sigma_{\rm Li}$ (S/cm) & $E_a$ (eV) & $D_{\rm Li}$ (cm$^2$/sec)\\ \hline
LGPS&$1.2\times10^{-2}$~\cite{Kamaya11}& 0.25~\cite{Kamaya11}&
\begin{tabular}{c}
$3\times10^{-8}$ (RT)~\cite{Kuhn13}\\
$6\times10^{-7}$ ($T=473$~K) 
\end{tabular}
\\ \hline
Li$_3$PS$_4$&$2\times10^{-4}$~\cite{Hayashi01}&0.35~\cite{Hayashi01}& 
\begin{tabular}{c}
$9\times10^{-10}$ (RT),~\cite{Stoffler18}\\
$^{*}2.9\times10^{-8}$ ($T=473$~K)
\end{tabular}
\\ \hline
Li$_7$P$_{3}$S$_{11}$ (glass) &
$0.91\times10^{-3}$~\cite{Onodera12}&0.44~\cite{Onodera12}&$5.7\times10^{-6}$ ($T=473$~K)~\cite{Mori15} \\ \hline
LLTO & 
$7\times10^{-5}$~\cite{Inaguma93}& 
\begin{tabular}{c}
0.4 ($T<400$~K),\\
0.15 ($T>400$~K)~\cite{Inaguma93}
\end{tabular} & 
\begin{tabular}{c}
$6.8\times10^{-6}$ ($T=600$~K), \\
$^{*}3.1\times10^{-6}$ ($T=473$~K)
\end{tabular}  
\end{tabular}
\end{ruledtabular}
\end{table*}

\section{Experimental Details}
A powder sample of $^{7}$Li-enriched LLTO was synthesized 
using a solid-state reaction technique for the stoichiometric mixtures of 
$^{7}$Li-enriched Li$_{2}$CO$_{3}$
(99.9\%, Cambridge Isotope Laboratories, Inc.), La$_{2}$O$_{3}$ 
(99.9\%, Miike Smelting Co., Ltd.),
and TiO$_{2}$ (99.9\%, Toho Titanium Co., Ltd.).
The quality of the obtained LLTO powder was confirmed by
powder x-ray diffraction measurements (SmartLab, Rigaku Corporation),
as shown in Fig.~S1 in the Supplemental Material~\cite{Supplement}.
Bulk single crystals of $^{7}$Li-enriched LLTO were grown using
a directional solidification method
under a growth condition recently reported for 
Li$_{x}$La$_{(1-x)/3}$NbO$_{3}$~\cite{Fujiwara16}.
The sample is stable in the air with no reaction to moisture.
The composition $x$ of the sample was determined as  
$x=0.13$ using an inductivity-coupled plasma optical emission spectrometer.
The obtained LLTO crystals were cut perpendicular to the growth direction 
to create a disk with a 20~mm diameter and 1.5~mm thickness.
The orientation of each disk was checked 
in 5~mm steps using an x-ray Laue diffractometer.
Six disks consisting of only a single domain with the same crystal orientation
were stacked and fixed using hydrogen-free glue (CYTOP). 
The total weight and volume of the stacked discs 
were 6.7~g and $\sim1$~cc, respectively.
Multiple scattering effect was not considered because
the fraction of multiple scattering was as small as 10\% estimated
by approximating a cylindrical sample shape ~\cite{Blech_65}. 
Resolution-limited Bragg peaks were confirmed
by neutron experiments,
which guarantees a bulk single grain of the stacked disk-shape crystals. 

All first-principles calculations were conducted 
within a generalized gradient approximation revised for solids, 
as developed by Perdew, Burke, and Ernzerhof~\cite{Perdew_97},
within the framework of the density functional theory~\cite{Hohenberg_64, Kohn_65}, 
using the plane-wave basis projector augmented wave (PAW) method~\cite{Blochl_94}.
For the PAW potentials, $2s$ and $2p$ electrons for O, $3p$, $3d$, and $4s$ electrons for Ti, 
and $4f$, $5s$, $5p$, $5d$, and $6s$ electrons for La were explicitly treated as valence electrons, 
with a plane-wave cutoff energy of 500~eV.
Unit cells of the tetragonal (La$_{0.5}$,Li$_{0.5}$)TiO$_3$ 
[(La$_1$,Li$_1$)Ti$_2$O$_6$ ]($P4/mmm$) structure were assumed and 
relaxed using a $7\times7\times4$ $k$-point mesh within the Brillouin zones 
generated using the Monkhorst-Pack scheme, 
and a convergence criterion for residual forces of 0.01~eV/\AA.
In this study, all first-principles molecular dynamics (FPMD) simulations were 
performed by Vienna ab initio simulation package (VASP) code~\cite{Kresse_96, Kresse_99}.
Supercells for FPMD simulations 
for the Li migration trajectory were constructed from $2\times2\times1$ unit cells 
of the relaxed tetragonal ($P4/mmm$) structure with one Li vacancy for Li migration. 
The total number of electrons for the FPMD supercell was adjusted to keep the system insulating. 
FPMD production runs were performed within the canonical ensemble 
(constant volume, temperature and number of atoms) using a Nos\'{e}-Hoover thermostat~\cite{Evans85}
at temperatures ($T$) of 1000~K, with a time step of 1~fs. 
The calculated temperature, ion-electron energy, and kinetic energy were stable
as a function of the simulation steps, demonsrating the stability of FPMD simulations 
(Figs.~S2-S4 in the Supplemental Material~\cite{Supplement}).
The structures were relaxed for 2~ps (2000 steps) to ensure that thermodynamic equilibrium 
had been reached before applying production runs of 
40~ps (40,000 steps) for the Li migration trajectory.

QENS measurements were conducted using a
time-of flight near-backscattering spectrometer, {\bf DNA} 
installed at the J-PARC MLF~\cite{Shibata15}.
The final neutron energy of $2.08$~meV was selected  
using Si (111) analyzers and a Bragg angle of 87.5$^{\circ}$.
A pulse shaping chopper rotating at 300~Hz with 3~cm slit 
provided an elastic resolution of 2.5~$\mu$eV and an energy transfer ($E$)
range of $-20<E<80 \mu$eV.
Neutron scattering data were obtained using the UTSUSEMI software~\cite{Utsusemi}.
The stacked single crystal sample was sealed in an aluminum can with air
and mounted to access the ($hhl$) horizontal scattering plane. 
There are some controversies in the structure of LLTO.
For the high Li-doped sample near $x=0.13$,
tetragonal cell with $a$($\sim 2^{1/2}a_{p}$)$\times a$($\sim2^{1/2} a_{p}$)$\times c$($\sim 2a_{p}$),~\cite{Varez_95}
where $a_{p}$ is a lattice parameter of cubic unit cell,
simple tetragonal cell with $a$($\sim a_{p}$)$\times a$($\sim a_{p}$)$\times c$($\sim 2a_{p}$),~\cite{Ibarra_00}
and orthogonal cell with $a$($\sim 2a_{p}$)$\times b$($\sim 2a_{p}$)$\times c$($\sim 2a_{p}$)~\cite{Chung_99, Inaguma_06}
have been proposed.
Throughout this paper, we label the momentum transfer 
in units of the reciprocal lattice vectors $a^{*}=b^{*}=1.147$~\AA$^{-1}$
and $c^{*}=0.811$~\AA$^{-1}$ in the orthogonal notation.
Figure~\ref{fig1}(b) shows three scan trajectories in the horizontal 
scattering plane ($hhl$).
The sample was rotated to make scattering trajectories 
nearly parallel to the three principal axes [110], [111], and [001]
below $Q<1.2$~\AA$^{-1}$.
The scattering intensity along the vertical direction was 
integrated for $-0.2<k<0.2$~r.l.u. in ($0,k,-k$).

\section{Results}
\begin{figure}
\includegraphics[keepaspectratio=true,width=8.5cm,clip]{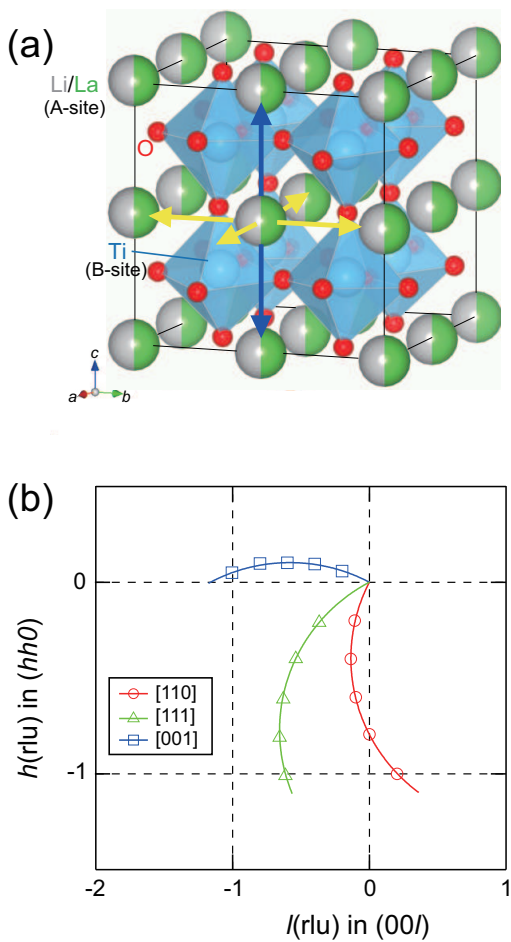}
\caption{\label{fig1} (color online)
(a) Crystal structure of  LLTO.
Arrows in (a) indicate jump vectors of Li$^+$ ions.
(b) Scan trajectory of neutron measurements.}
\end{figure}

\begin{figure}
\includegraphics[keepaspectratio=true,width=8.5cm,clip]{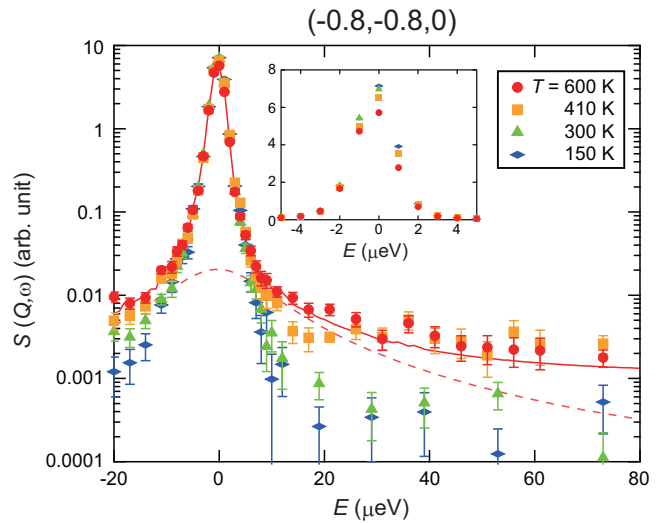}
\caption{\label{fig2} (color online)
QENS spectra at ($-0.8,-0.8,0$) for $^{7}$LLTO
measured at $T=150, 300, 410$, and 600~K.
The inset shows enlarged data near $E=0$ in a linear scale.
Scattering intensities for $-0.9<h<-0.7$~r.l.u. in ($hh0$)
are averaged.
The solid line indicates a fit with the sum of Lorentzian
and delta functions, and a flat background.
The $T=150$~K data were convoluted to the delta and Lorentzian 
functions as an instrumental resolution.
The dotted line represents the Lorentzian function 
for the fitting to the 600~K data.
}
\end{figure}

The QENS spectra at (-0.8,-0.8,0) at various $T$'s are plotted
in the logarithmic scale in Fig.~\ref{fig2}.
Because Li$^+$ ions are expected to be immobile at $T=150$~K,
we assume that the spectrum at $T=150$~K corresponds to a delta function
convoluted with the instrumental resolution.
A similar resolution-limited spectrum is also obtained at $T=300$~K, 
suggesting that the Li$^+$ ions are still immobile even at $T=300$~K
within the energy or time resolution of the current setup 
(2.5~$\mu$eV$\sim0.6$~GHz$\sim0.5$~ns).
At $T\ge410~$K, 
the intensity of the elastic peak decreases with increasing $T$ (see the inset of Fig.~\ref{fig2}),
whereas the scattering intensities for $|E|>4$~$\mu$eV are enhanced with $T$. 
This suggests the dynamic nature of Li$^+$ ions at $T\ge410$~K.
These spectra were fitted with the sum of a Lorentzian and a delta function 
convoluted with the resolution function estimated from the $T=150$~K data
in addition to a flat background.
Although the QENS component, i.e., the Lorentzian component,
is 1000 times weaker than the elastic peak, 
the QENS component, represented by a dotted line in Fig.~\ref{fig2}, 
is clearly extracted, mainly owing to the low background of the DNA spectrometer.
Note that the half width at half maximum ($\mathit{\Gamma}$)
of a Lorentz function was $10\pm2$~$\mu$eV,
which is comparable to $\mathit{\Gamma}$ for 
a sulfide-based Li ionic conductor
Li$_{7}$P$_{3}$S$_{11}$ obtained using {\bf DNA}~\cite{Mori14}.

\begin{figure}
\includegraphics[keepaspectratio=true,width=8.5cm,clip]{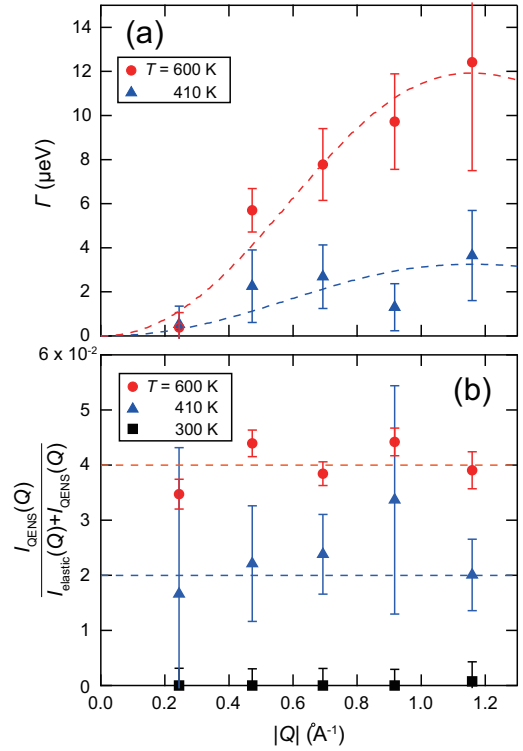}
\caption{\label{fig3} (color online)
$Q$ dependences of (a) Lorentzian line width $\Gamma$ and
(b) QENS to total intensity ratio ($V_{\rm QENS}$)
for the [110] scan trajectory, as described in the text. 
$\Gamma$ is determined from the data averaged
over $\Delta h=\pm0.1$~r.l.u. in ($hh0$)
at approximately $h=-1.0$, -0.8, -0.6, -0.4, and -0.2~r.l.u.
The magnitude of a wave vector from the origin $Q$
is calculated at each point and $\Gamma$ is plotted against $Q$.
}
\end{figure}

From the fits at $h=-1.0$, -0.8, -0.6, -0.4, -0.2 (rlu) 
along the scan trajectory of [110] [Fig.~\ref{fig1}(b)],
the $Q$ dependence of $\mathit{\Gamma}$
was obtained at $T=410$ and 600~K [Fig.~\ref{fig3}(a)], 
where $Q$ is the magnitude of a wave vector from the origin.
Because $\mathit{\Gamma}$ increases quadratically with $Q$ and saturates at a high $Q$, 
the obtained result suggests a translational jump diffusion of Li$^+$ 
for $T\ge410~$K.

In the Chudley-Elliot (CE) model for 
the translational jump diffusion in Bravais lattices,
the scattering function $S(\mbox{\boldmath $Q$},\omega)$ is given by~\cite{Hempelmann}
\begin{equation}
S(\mbox{\boldmath $Q$},\omega)=\frac{\mathrm{exp}(-2W)}{\pi}\frac{\mathit{\Gamma}(\mbox{\boldmath $Q$})}{\omega^2+\mathit{\Gamma}(\mbox{\boldmath $Q$})^{2}},
\label{eq1}
\end{equation}
where $\mathrm{exp}(-2W)$ is the Debye-Waller factor.
Here, $\mathit{\Gamma}(\mbox{\boldmath $Q$})$ is written as
\begin{equation}
\mathit{\Gamma}(\mbox{\boldmath $Q$})=\frac{\hbar}{z\mathit{\tau}} \sum_{j}^{z} [1-\mathrm{exp}(-i\mbox{\boldmath $Q$}\cdot\mbox{\boldmath $l$}_{j})],
\label{eq2}
\end{equation}
where $\mathit{\tau}$ is the residence time, 
$\mbox{\boldmath $l$}_{j}$ is the jump vector, 
and $z$ is the number of sites.
Because $[1-\mathrm{exp}(-i\mbox{\boldmath $Q$}\cdot\mbox{\boldmath $l$}_{j})]=0$ 
when $\mbox{\boldmath $Q$}\bot\mbox{\boldmath $l$}_{j}$,
$\mathit{\Gamma}(\mbox{\boldmath $Q$})$ for the [110] direction reflects 
the Li$^+$ jumps in the $ab$ plane.

\begin{figure}
\includegraphics[keepaspectratio=true,width=8.5cm,clip]{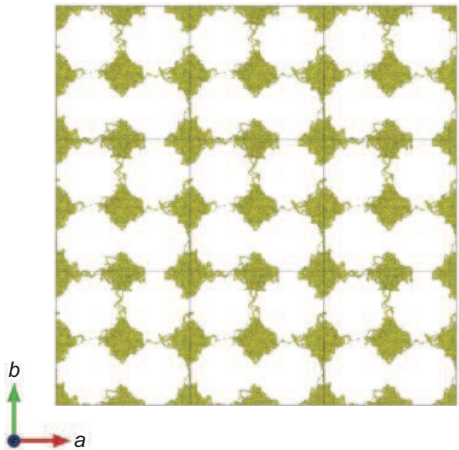}
\caption{\label{fig4} (color online)
Trajectories of Li$^+$ ions in the $ab$ plane of the La-poor layer
simulated by MD using first-principles modeling for LLTO.
The yellow traces show trajectories of Li$^+$ ions.
}
\end{figure}

To infer the candidates of the $\mbox{\boldmath $l$}_{j}$ in the $ab$ plane,
we conducted MD simulations using first-principles calculations of LLTO.
The simulated trajectories of Li$^+$ ions
in the La-poor (La2) layer on the $ab$ plane are shown in Fig.~\ref{fig4}.
For simplicity, we also assumed the absence of La ions 
at the La2 site in the simulations.
Li$^+$ ions move around the $A$ sites, and 
the trajectories form a square shape with edges toward 
the $a$ or $b$ directions.
The square trajectories are consistent with the distribution of Li$^+$ ions
determined from reverse Monte Carlo simulations based on the
x-ray and neutron diffraction data~\cite{Ohara10, Mori14}.
Note that some of the Li$^+$ ions jump to the neighboring $A$ site 
through the bottle neck structure,
suggesting that the jump toward the vacant $A$ site 
is the key pathway of Li$^+$ diffusion.

Based on the information obtained from the MD calculations, 
Li$^+$ ions are found to jump along the [100] and [010] directions in the $ab$ plane. 
Therefore, Eq.~(\ref{eq2}) is rewritten as
\begin{equation}
\mathit{\Gamma}(\mbox{\boldmath $Q$})=\frac{\hbar}{2\mathit{\tau}_{ab}} [2-\mathrm{cos}(\frac{Q_{x}l_{ab}}{\sqrt{2}})-\mathrm{cos}(\frac{Q_{y}l_{ab}}{\sqrt{2}})]. 
\label{eq3}
\end{equation}
The observed $\mathit{\Gamma}(\mbox{\boldmath $Q$})$ 
for the [110] direction are reproduced well 
by this model with $\mathit{\tau}_{ab}=0.11\pm0.01$~ns and $l_{ab}=3.9\pm0.4$~\AA\,
as indicated by the broken lines in Fig.~\ref{fig3}(a).
Note that the obtained $l_{ab}$ is equivalent to the $a$-axis length 
of the cubic perovskite ($a=3.87~$\AA) within the error bar. 
This is consistent with the result of the MD calculations, 
in which Li$^+$ ions jump to the neighboring $A$ site
through the bottle neck.

For a small $Q$ parallel to the jump vector, 
Eq.~(\ref{eq2}) is converted into the following: 
\begin{equation}
\mathit{\Gamma}(Q)=\hbar \frac{{l_{j}}^2}{2\mathit{\mathit{\tau}}}Q^{2}=\hbar D_{\rm Li}Q^{2}, 
\label{eq4}
\end{equation}
where $D_{\rm Li}$ is a self-diffusion coefficient of Li$^+$ ions.
Thus, $D_{\rm Li}$ is estimated as $6.8\pm0.5\times10^{-6}$~cm$^{2}$/s at $T=600$~K
directly from ${l_{j}}^2/2\mathit{\tau}$.
Because the jump distance $l$ is roughly independent of $T$,
$\mathit{\tau}$ and $D_{\rm Li}$ are easily estimated at $T=410$~K (see Table~\ref{tb2}).
Inaguma {\it et al.} reported the thermal activation energy ($E_{a}$) 
of 0.15~eV at above 400~K from the bulk conductivity measurements~\cite{Inaguma93}.
The decrease in $D_{\rm Li}$ is in good agreement with 
the Arrhenius law with $E_{a}$ of 0.15~eV.
It is interesting to compare the present $D_{\rm Li}$ with that for Li$_{7}$P$_{3}$S$_{11}$
estimated from the QENS data obtained with {\bf DNA}:
that is, for Li$_{7}$P$_{3}$S$_{11}$, $D_{\rm Li}=5.7\times10^{-6}$~cm$^{2}$/s 
at $T=473$~K~\cite{Mori15}.
For LLTO, 
$D_{\rm Li}$ at $T=473$~K is estimated from 
$D_{\rm Li}$ at 600~K and $E_{a}=0.15$~eV
as $3.1\pm0.3\times10^{-6}$~cm$^{2}$/s, 
which is close to that for Li$_{7}$P$_{3}$S$_{11}$ (Table~\ref{tb1}). 
This means that a fast Li$^+$ diffusion is achieved 
even in oxide-based Li$^+$ ionic conductors, 
as well as in the sulfide-based conductors, 
from a microscopic perspective.

In addition to $\mathit{\Gamma}$($Q$),
a fraction of the QENS component to the total intensity 
($V_{\rm QENS}$), defined as
\begin{equation}
V_{\rm QENS}=\frac{I_\mathrm{QENS}(Q)}{I_\mathrm{elastic}(Q)+I_\mathrm{QENS}(Q)},
\label{eq5}
\end{equation}
provides useful information on the Li$^+$ dynamics.
Here, $I_\mathrm{QENS}$($Q$) and $I_\mathrm{elastic}$($Q$) are 
the integrated intensities of the QENS and elastic signal, respectively.
For the translational diffusion, 
because only the Debye-Waller factor depends on $Q$
in both $I_\mathrm{QENS}$($Q$) and $I_\mathrm{elastic}$($Q$),
such $Q$ dependence is canceled out.
Therefore, $V_{\rm QENS}$ is also independent of $Q$
and is a good indicator for the number of mobile Li$^+$ ions.
In fact, $V_{\rm QENS}$ is roughly independent of $Q$, as shown in Fig.~\ref{fig3}(b), 
which confirms the self-diffusive nature of the observed QENS signal
at $T=600$~K and 410~K.
The mobile Li$^+$ ion fraction can be obtained by multiplying 
$V_{\rm QENS}$ by the ratio of the incoherent scattering cross section 
$\sigma_{\rm inc}$ of $^{7}$Li to the sum of $\sigma_{\rm inc}$ 
from all elements contained in LLTO, as summarized in Table~\ref{tb2}.
The reduction in the fraction of mobile Li$^+$ ions 
at lower $T$ indicates the decrease 
in the number of mobile Li$^+$ ions detectable in the time window of 
the current experimental setup.

\begin{table*}
\caption{\label{tb2}Jump distance $l$, residence time $\mathit{\tau}$, diffusion coefficient $D_{\rm Li}$, 
ratio of mobile Li$^+$ ions, and the number of mobile Li$^+$ ions $N$ for LLTO ($x=0.13$).}
\begin{ruledtabular}
\begin{tabular}{c|c|c|c|c|c|c}
Direction & $T$(K) & $l$ (\AA) & $\mathit{\tau}$ (ns) & $D_{\rm Li}$ ($10^{-6}$ cm$^{2}$/s) & Ratio of mobile Li$^+$ ions(\%) & $N$ ($10^{27}$ m$^{-3}$) \\ \hline
[110] & 600 & $3.9\pm0.4$ & $0.11\pm0.01$ & $6.8\pm0.5$ & $50\pm1$ & $3.35\pm0.07$ \\ \relax
[110] & 410 & 3.9 fix & $0.42\pm0.09$ & $1.9\pm0.1$ & $25\pm9$ & $1.7\pm0.6$ \\ \relax
[001] & 600 & $3.5\pm0.3$ & $0.14\pm0.01$ & $4.4\pm0.3$ & $24\pm2$ & $1.6\pm0.2$
\end{tabular}
\end{ruledtabular}
\end{table*}

\begin{figure}
\includegraphics[keepaspectratio=true,width=8.5cm,clip]{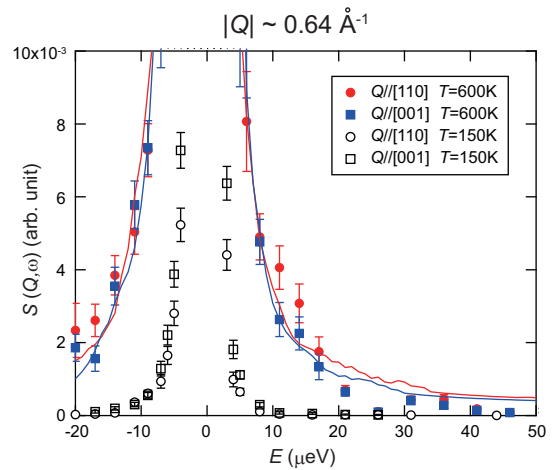}
\caption{\label{fig5} (color online)
In-plane ([110]) and out-of-plane ([001]) QENS 
spectra at $|Q|\sim0.64$~\AA$^{-1}$
measured at $T=150$ and 600~K.
To compare the intensity for different directions,
neutron transmissions were corrected 
assuming a rectangular shape of the sample.
The in-plane spectra were measured at ($-0.55,-0.55,-0.11$), 
whereas the out-of-plane spectra were obtained at ($0.09,0.09,-0.80$).
The solid lines represent the fits 
convolved with $T=150$~K data as instrumental resolution.
}
\end{figure}

Next, to determine the anisotropy of Li$^+$ dynamics,
the in-plane ([110])-QENS spectrum is compared with the
out-of-plane ([001])-QENS spectrum
at the same $|Q|\sim0.64$~\AA$^{-1}$.
Figure~\ref{fig5} shows the spectra measured at $T=150$ and 600~K.
At $T=600$~K, the [110]-QENS component is similar to 
the [001]-QENS component,
indicating a nearly three-dimensional motion of Li$^+$ ions.
In fact, the two QENS spectra are well reproduced by the common 
$\mathit{\Gamma}$ of $7\pm2$~$\mu$eV, 
whereas the QENS intensity of the in-plane motion is slightly larger 
than that of the out-of-plane motion.

\begin{figure}
\includegraphics[keepaspectratio=true,width=8.5cm,clip]{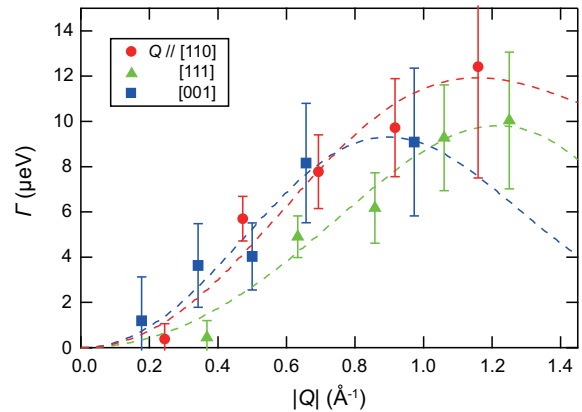}
\caption{\label{fig6} (color online)
(a) $Q$ dependences of HWHM $\Gamma$ of the
Lorentzian component along
three principal axes [110], [111], and [001] 
measured at $T=600$~K.
}
\end{figure}

Figure~\ref{fig6} shows the $\mathit{\Gamma}(\mbox{\boldmath $Q$})$ 
curves at 600~K along the [110], [001], and [111] directions.
For the [001] direction,
$\mathit{\Gamma}(\mbox{\boldmath $Q$})$ is described as
\begin{equation}
\frac{\hbar}{\mathit{\tau}_{c}} [1-\mathrm{cos}(Q_{z}l_{c})],
\label{eq6}
\end{equation}
with $l_{c}=3.5$~\AA\, and $\tau_{c}=0.14$~ns (Table~\ref{tb2}).
The jump length along the [001] direction ($l_{c}$) is close to 
the $a$-axis length of the cubic cell,
suggesting the jumps between the $A$-sites of the 
neighboring Li/La planes [Fig.~\ref{fig1}(a)].
Because $\tau_{[001]}$ is slightly larger than $\tau_{[110]}$, 
$D_{\rm Li[001]}<D_{\rm Li[110]}$. 
The $\mathit{\Gamma}$($\mbox{\boldmath $Q$}$) curve along the [111] direction
is explained well by the coexistence of both jumps in the $ab$ plane
and along the $c$ axis using the obtained QENS parameters 
for each direction.

\section{Discussion}
Returning to the Nernst-Einstein equation,
$D_{\rm Li}$ is related to the ionic conductivity $\sigma_{\rm Li}$ as
\begin{equation}
\sigma=e^{2}ND/k_{B}T,
\label{NE}
\end{equation}
where $e$ is the elementary charge, 
$N$ is the density of mobile Li$^+$ ions, and
$k_{\rm B}$ denotes the Boltzmann constant.
$N$ is estimated as $3.35\pm0.07\times10^{27}$~m$^{-3}$ 
from the product of the mobile Li$^+$ ion fraction
and the number of Li$^+$ ions ($3x$) per cubic perovskite unit cell
$6.7\times10^{27}$ m$^{-3}$
(Table~\ref{tb2}).
The microscopic $\sigma_{\rm Li}$
(hereafter denoted as $\sigma_{\rm QENS}$)
at $T=600$~K is calculated as $0.070\pm0.006$~S/cm
and $0.02\pm0.01$~S/cm for the in-plane and out-of-plane directions, 
respectively.
The spatial average of the $\sigma_{\rm QENS}$ 
$0.055\pm0.005$~S/cm
agrees with that of $\sigma_{Li}\sim0.1$~S/cm at $T=600$~K
reported from bulk conductivity measurements
for LLTO with slightly lower Li content ($x=0.11$)~\cite{Inaguma93}.

Table~\ref{tb1} summarizes $\sigma_{Li}$, $E_{a}$, and $D_{\rm Li}$
for the sulfide-based Li$^+$ ion conductors and LLTO.
The $\sigma_{Li}$ of LLTO is one or two orders of magnitude lower than
that of the sulfur-based ones with relatively large $E_{a}$.
The $D_{\rm Li}$ for LGPS and Li$_7$P$_{3}$S$_{11}$ are the reported values obtained from pulsed
field gradient nuclear magnetic resonance measurements (PFG-NMR),
and they are considerably smaller than those of 
LLTO and Li$_7$P$_{3}$S$_{11}$ despite their higher conductivity. 
Presumably, this is
because QENS capture the dynamics of the smallest Li dynamic path,
whereas PFG-NMR detects dynamics with longer length scale.
Comparing the $D_{\rm Li}$ measured by QENS between LLTO and Li$_7$P$_{3}$S$_{11}$,
the fact that $D_{\rm Li}$ for LLTO is comparable to that for Li$_{7}$P$_{3}$S$_{11}$,
despite the difference in $\sigma_{Li}$, leads to the following question:
what is the predominant factor for the difference in $\sigma_{\rm Li}$
between LLTO and Li$_{7}$P$_{3}$S$_{11}$?
The $V_{\rm QENS}$ for Li$_{7}$P$_{3}$S$_{11}$ is 0.23 at $T=473$~K~\cite{private}
and the fraction of mobile Li$^+$ ions is 23\%,
which is similar to that for LLTO (Table~\ref{tb2}).
Then, $N$ is obtained as $3.88\times10^{27}$~m$^{-3}$ 
by multiplying a Li ion density $1.69\times10^{28}$ m$^{-3}$
for Li$_{7}$P$_{3}$S$_{11}$.
$N$ for LLTO at $T=473$~K is $2.2\times10^{27}$~m$^{-3}$,
estimated from a linear interpolation between $T=410$ and 600~K,
which is 1.7 times smaller than that of Li$_{7}$P$_{3}$S$_{11}$.
Consequently, 
$\sigma_{\rm QENS}$ for Li$_{7}$P$_{3}$S$_{11}$ is expected to be 
3.2 times larger than that for LLTO
according to the Nernst-Einstein equation.
This is the most reasonable explanation for 
the difference in $\sigma_{\rm Li}$
between LLTO and Li$_{7}$P$_{3}$S$_{11}$ at the microscopic scale.

Since the discovery of fast Li$^+$ ionic conduction 
in sulfide-based Li$^+$ ionic conductors~\cite{Kanno_00},
sulfide ions are considered to play a significant role in enhancing 
the mobility of Li$^+$ ions through a large polarization. 
However, the fact that $D_{\rm Li}$ for LLTO is comparable to that of Li$_{7}$P$_{3}$S$_{11}$
indicates the possibility of better oxide-based Li$^+$ ionic conductors
by seeking materials with a larger Li density in the unit cell.
Regarding the microscopic mechanism of fast Li$^+$ diffusion,
a coupling between the rotational motion of the PS$_{4}^{3-}$ tetrahedra
and $\sigma_{\rm Li}$ was indicated 
in the sulfide-based Li$^+$ ionic conductors~\cite{Smith20}.
In LLTO, structural instabilities of the tilting modes of TiO$_{6}$ tetrahedra
are suggested from first-principles calculations~\cite{Moriwake15}.
Further studies on the dynamics of the host lattice of Li$^+$ ionic conductors  
will be required to understand the mechanism of fast Li$^+$ ionic conduction in solids.

\section{Conclusion}
We studied the dynamics of Li$^+$ ions in a solid-state electrolyte 
LLTO ($x=0.13$) using QENS and first-principles MD simulations.
We observed clear QENS signals  along the three principal axes 
[110], [111], and [001] at $T=600$~K 
by using the large $^{7}$Li-enriched single crystal.
Directional information of the Li$^+$ jump, such as the residence time 
and jump vector determined at the microscopic scale,
reveals the dynamical conduction paths of Li$^+$ ions:
jumps of Li$^+$ ions to the neighboring $A$ site
through the bottle neck structure,
which are supported by the first-principles MD simulations.
The self-diffusion coefficients of Li$^+$ ions are found to be quasi-isotropic,
suggesting the nearly three-dimensional diffusion of Li$^+$ ions.
Furthermore,
the estimated self-diffusion coefficients of Li$^+$ are comparable 
to those of the sulfide-based Li$^+$ ion conductor Li$_7$P$_3$S$_{11}$, 
although the ionic conductivity of LLTO is 10 times smaller 
than that of Li$_7$P$_3$S$_{11}$.
This microscopic information on Li$^+$ diffusion can provide 
a way to increase Li$^+$ ionic conductivity in oxide-based solid electrolytes.

\section{Acknowledgement}
The neutron experiments were performed with the approval of J-PARC MLF 
(No. 2019A0306).
We are grateful to K. Mori, M. Kofu, and J. Sugiyama for helpful discussions.
We also thank M. Fujita and Y. Ikeda for their help with crystal alignment 
using a x-ray Laue diffractometer.
The crystal structures in Fig.~\ref{fig1}(a) are produced by VESTA software\cite{Momma_11}.

%

\end{document}


\preprint{}

\title{Supplemental material for\\Microscopic dynamics of lithium diffusion 
in single crystal of solid-state electrolyte La$_{2/3-x}$Li$_{3x}$TiO$_{3}$ ($x=0.13$)
studied by quasi-elastic neutron scattering
}

\author{Masato~Matsuura$^{*}$}
  \affiliation{Neutron Science and Technology Center, Comprehensive Research Organization for Science and Society (CROSS), Tokai, Ibaraki 319-1106, Japan}
  \email{m_matsuura@cross.or.jp}
\author{Yasuyuki Fujiwara}
  \affiliation{Faculty of Engineering, Shinshu University, Nagano 380-8553, Japan}
\author{Hiroki Moriwake}
  \affiliation{Nanostructures Research Laboratory, Japan Fine Ceramics Center, Nagoya 456-8587, Japan}
\author{Koji Ohara}
  \affiliation{Research and Utilization Division, Japan Synchrotron Radiation Research Institute (SPring-8/JASRI),  Sayo, Hyogo 679-5198, Japan}
\author{Yukinobu Kawakita}
  \affiliation{Materials and Life Science Division, J-PARC Center, JAEA, Tokai, Ibaraki 319-1195, Japan}
\date{\today}

\maketitle

\section{Supplementary Material}

\begin{figure}
\includegraphics[keepaspectratio=true,width=10cm,clip]{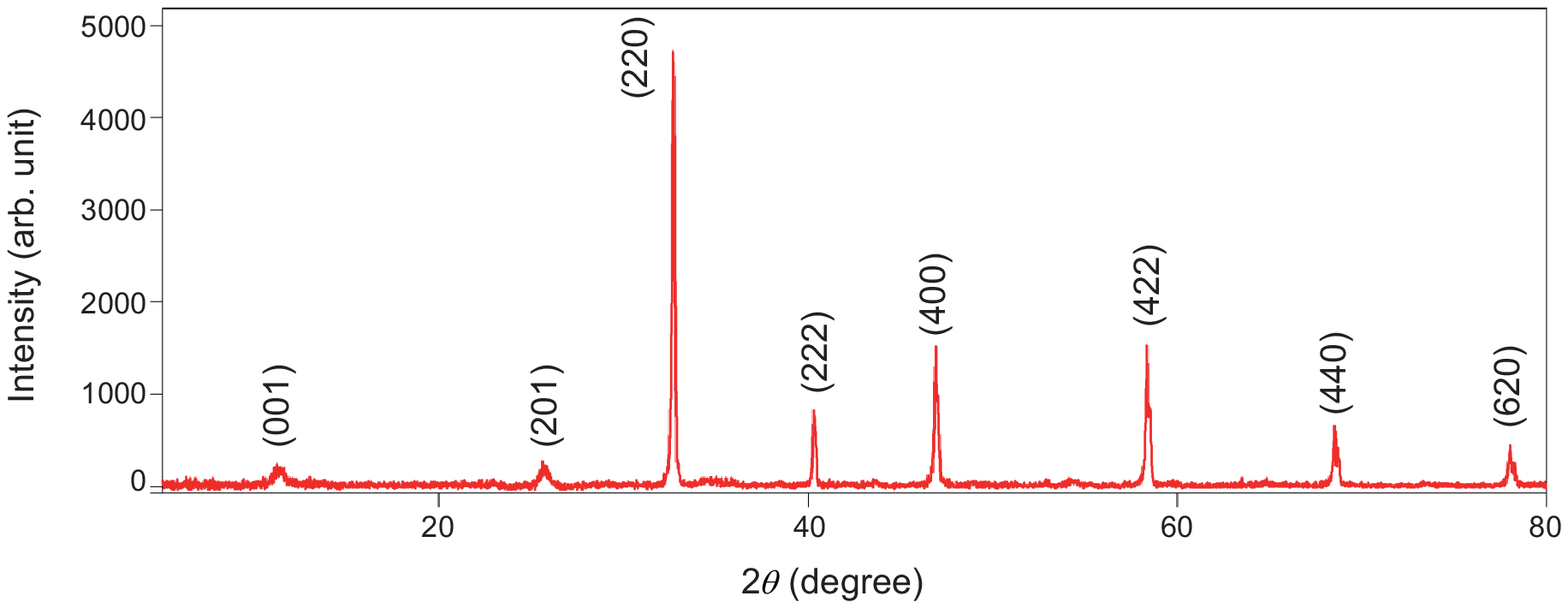}
\caption{\label{fig1} (color online)
Powder X-ray diffraction pattern of LLTO ($x=0.13$) used in this study. }
\end{figure}

Figure~\ref{fig1} shows X-ray powder diffraction pattern of 
La$_{2/3-x}$Li$_{3x}$TiO$_{3}$ (LLTO) with $x=0.13$ used in this study.
All reflections were indexed with Orthogonal cell [space group: $Cmmm$,
$a$($\sim 2a_{p}$)$\times b$($\sim 2a_{p}$)$\times c$($\sim 2a_{p}$)~\cite{Chung_99, Inaguma_06}],
indicating no impurity phase in the sample.

\begin{figure}
\includegraphics[keepaspectratio=true,width=9cm,clip]{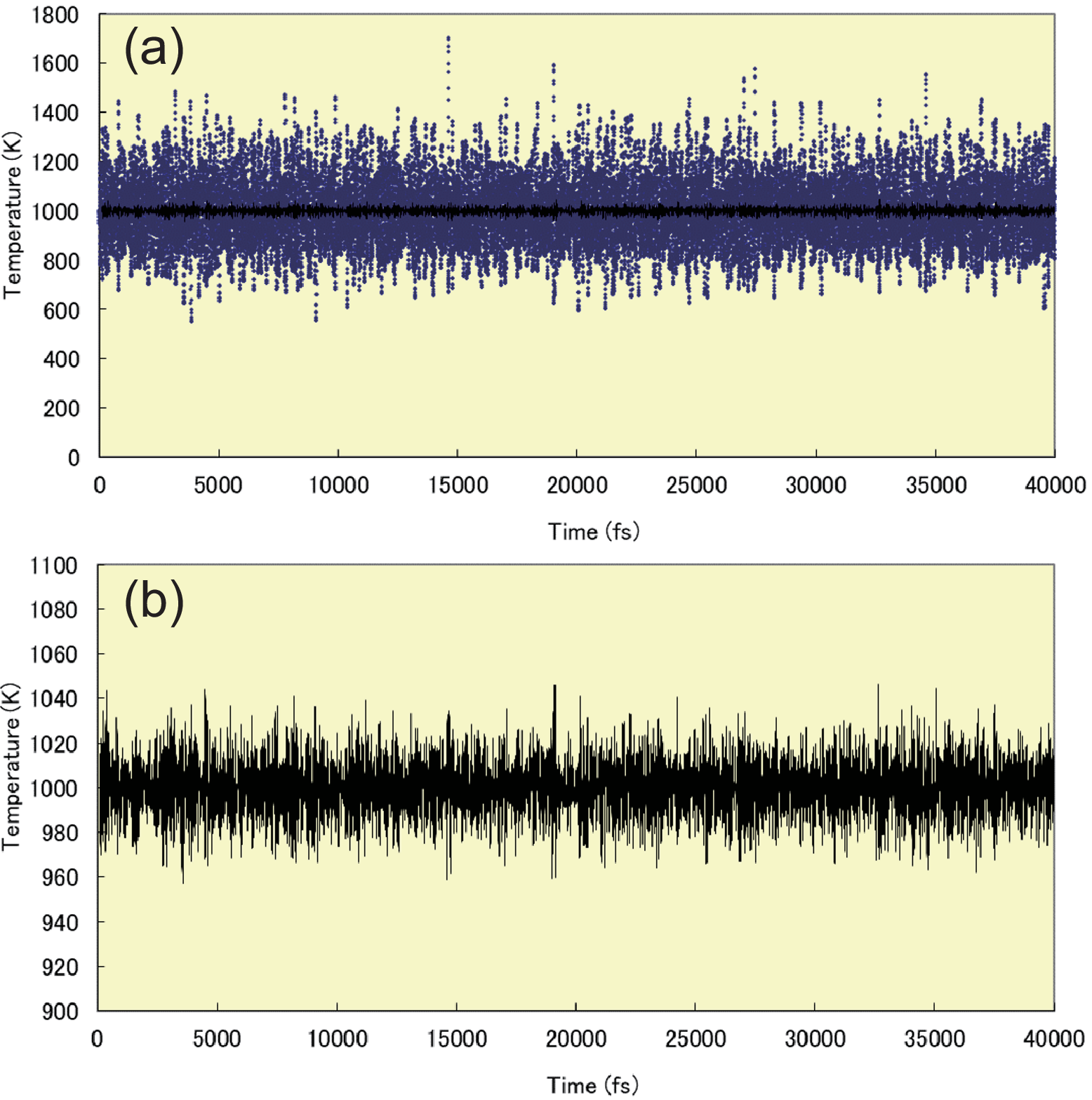}
\caption{\label{fig2} (color online)
(a) Temperature drift in the production run. 
The blue dots indicate calculated temperature, 
whereas the black line indicates average over 100 fs (step). 
(b) Enlarged plot of average temperature. }
\end{figure}

\begin{figure}
\includegraphics[keepaspectratio=true,width=9cm,clip]{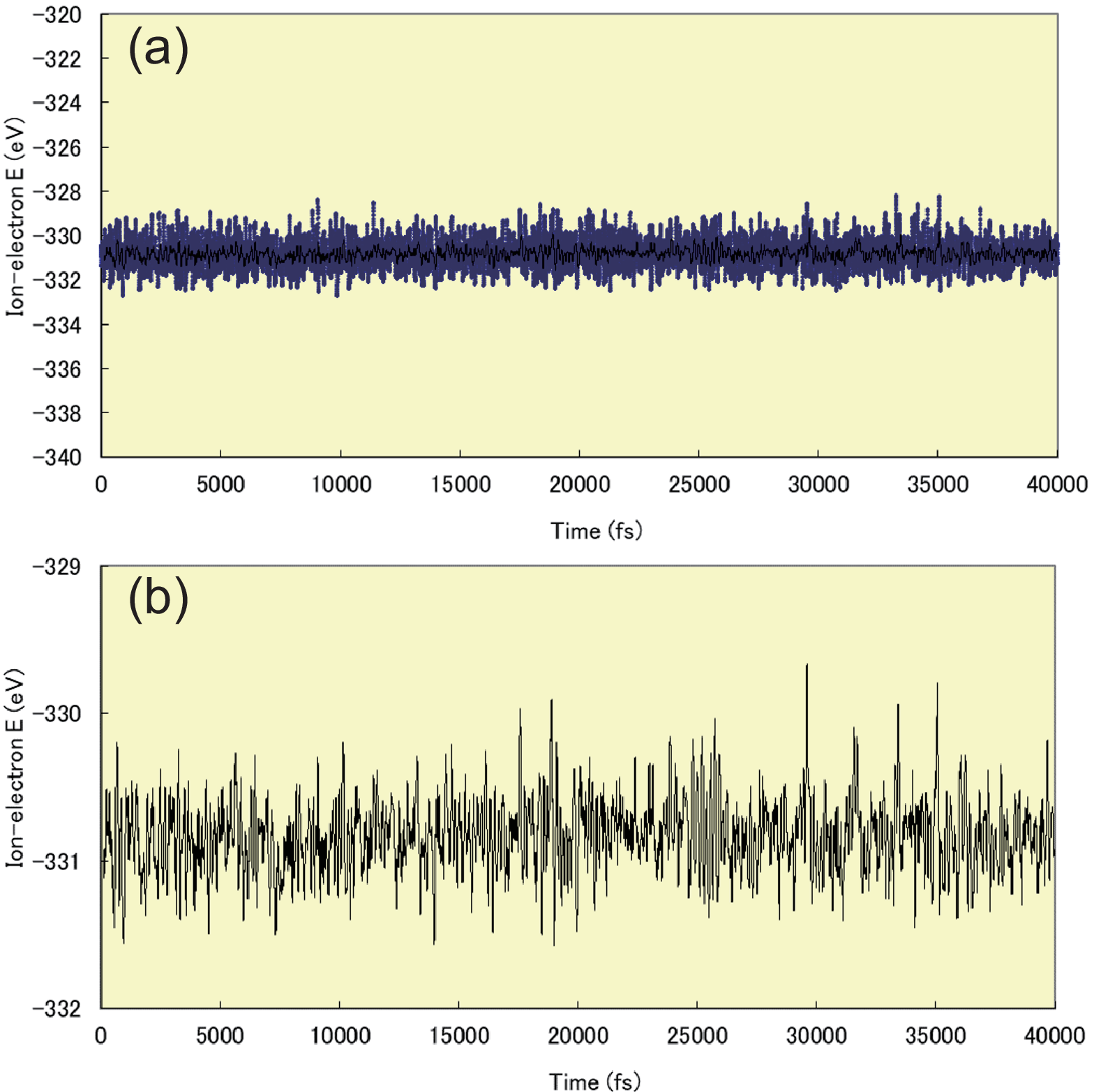}
\caption{\label{fig3} (color online)
(a) Ion-electron energy drift in the production run. 
The blue dots indicate calculated ion-electron energy, 
whereas the black line indicates average over 100 fs (step). 
(b) Enlarged plot of average ion-electron energy. }
\end{figure}

\begin{figure}
\includegraphics[keepaspectratio=true,width=9cm,clip]{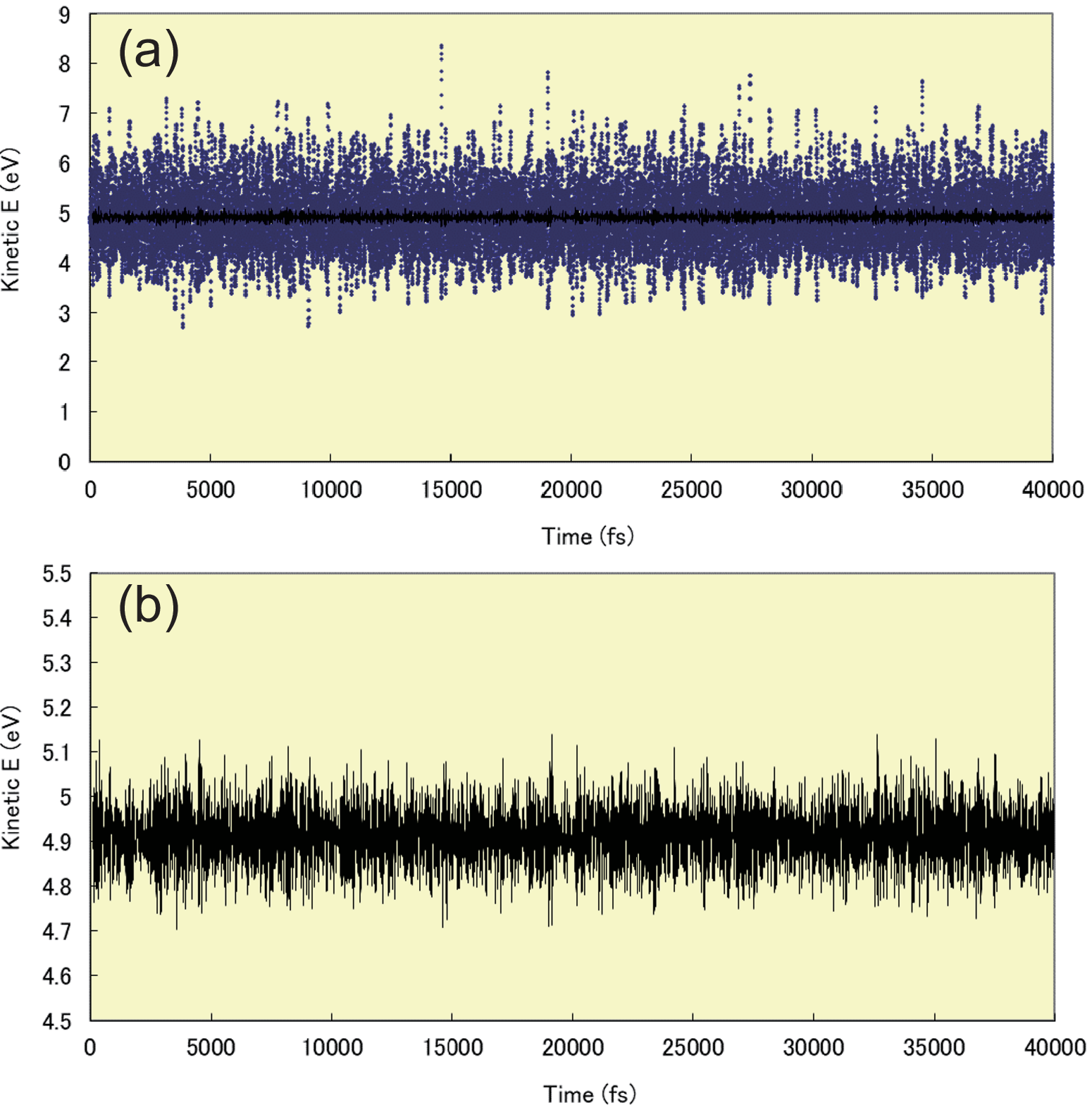}
\caption{\label{fig4} (color online)
(a) Kinetic energy drift in the production run. 
The blue dots indicate calculated ion-electron energy, 
whereas the black line indicates average over 100 fs (step). 
(b) Enlarged plot of average kinetic energy. }
\end{figure}

To demonstate the stability of first-principle molecular dynamics (FPMD) simulations
in this study, here, we present certain plots of energy and temperature stability
as a function of the simulation steps (40000).
Drifts in the calculated temperature, ion-electron energy, and 
kinetic energy are shown in Fig.~\ref{fig2}, Fig.~\ref{fig3}, and Fig.~\ref{fig4},
respectively.
No significant drift was observed in the calculated temperature,
ion-electron energy, and kinetic energy, confirming the 
stability of our FPMD simulations.
Furthermore, the averaged temperature, ion-electron energy, and kinetic energy were
controlled at $1000\pm20$~K, $330.8\pm0.5$~eV, and $4.91\pm0.1$~eV,
respectively.

\bibliography{LLTO_QENS}